\documentclass[5p]{elsarticle}
\usepackage{amsmath,amssymb,graphicx,url,hyperref,breakurl}

\title{Testing for Detailed Balance in a Financial Market}

\author[fiu]{H.R.~Fiebig\corref{cor1}}
\author[fiu]{D.P.~Musgrove}

\address[fiu]{Department of Physics, Florida International University, Miami, Florida 33199, USA}

\cortext[cor1]{Corresponding author}

\begin{document}

\begin{abstract}
We test a historical price time series in a financial market (the NASDAQ~100 index) for
a statistical property known as detailed balance. The presence of detailed balance would
imply that the market can be modeled by a stochastic process based on a Markov chain,
thus leading to equilibrium. In economic terms, a positive outcome of the test would support
the efficient market hypothesis, a cornerstone of neo-classical economic theory.
In contrast to the usage in prevalent economic theory the term equilibrium here is tied
to the returns, rather than the price time series.
The test is based on an action functional $S$ constructed from the elements of the
detailed balance condition and the historical data set, and then analyzing $S$ by means of
simulated annealing.
Checks are performed to verify the validity of the analysis method.
We discuss the outcome of this analysis.
\end{abstract}


\maketitle

\section{\label{sec:intro}Introduction}

Much of contemporary economic theory is dominated by the neo-classical paradigm of an efficient market.
For definiteness, considering financial markets, this implies that
the market participants (traders), have immediate and complete access to market
information, like stock prices, sales volumes etc, and engage in rational behavior
(trade decisions), aiming to maximize their self interest.
It is argued that this situation will then lead to some kind of equilibrium state
of the market, where the actual price of a financial instrument reflects its real
market value at all times \cite{Smith:Adam}.
In this context, equilibrium here means that the price fluctuates stochastically
about some average value. The fluctuations, caused by the interactions of many traders,
will influence the price only on a short time scale.

In an alternative scenario a market may be in an off-equilibrium state \cite{McCauley:2009}.
This paradigm would allow for dramatic price changes, catastrophic crashes in particular.
A signature feature is a power law behavior of the returns distribution for extreme events,
much resembling the Gutenberg-Richter law for earthquakes.
This suggests that methods from the field of critical systems \cite{Sornette:2000}
might be fruitful.  In the context of a financial time series,
this situation can also emerge from a self-organized critical state \cite{Dupoyet2010107,Dupoyet20113120}.

We are here interested in the question of whether a real market does reveal signs
if equilibrium. Our approach is entirely based on the analysis of empirical, historical, data.
For this purpose an operational definition of `equilibrium' in a financial time series, the subject
of our investigation, is required. We will motivate our choice (somewhat different from customary use)
in Sect.~\ref{sec:two}, and then define the criterion for the test in Sect.~\ref{sec:three}.
We discuss the numerical implementation and outcome of the test in Sect.~\ref{sec:four},
closing with Sect.~\ref{sec:conclusion}, which contains the conclusion.

\section{\label{sec:two}Motivation}

For the sole purpose of motivating our criterion, we briefly reflect on the well-known Metropolis
algorithm \cite{metropolis1953} which is an standard tool in numerical simulation for
generating sets of random numbers according to a given
probability density function. The algorithm is one way of producing a Markov chain of numbers
\cite{feller1,feller2}, say
\begin{equation}
\ldots \leftarrow r(i+1) \leftarrow r(i) \leftarrow r(i-1) \leftarrow \ldots\,,
\label{Mchain}\end{equation}
where a (real) value $r(i)$, at simulation `time' counter $i$, is generated from the preceding one
through a stochastic process. The latter involves a conditional probability density
function, say $W(r^\prime\leftarrow r)$. In the Metropolis algorithm it is
constructed from a base probability distribution function, say $w(r)$, by creating a trial
value $r^\prime$ and then accepting or rejecting it as the next value $r^\prime\leftarrow r$
in the chain as determined by $W(r^\prime\leftarrow r)$.
The ensuing Markov chain will eventually, in the limit
$\infty \leftarrow i$, produce values $r=r(i)$ distributed according to the base probability
density function $w(r)$. 
Having converged to $w(r)$, the chain is said to have reached `equilibrium'.
A property known as detailed balance
\begin{equation}
W(r^\prime\leftarrow r) w(r) = W(r\leftarrow r^\prime) w(r^\prime)
\label{Dbalance}\end{equation}
is a sufficient condition for the chain to reach equilibrium.
The Metropolis algorithm makes a very specific choice for $W(r^\prime\leftarrow r)$.
However, we will make no use of it until later in Sect.~\ref{sec:four} when we verify
and discuss our results.

We here adopt the detailed balance condition (\ref{Dbalance}) as the criterion to be tested
for in a historical data set.
Strictly speaking, it is `only' a sufficient condition for equilibrium. On the other hand, the
notion  of equilibrium is rather fuzzy as used in an economic context. Detailed balance has
the advantage of providing us with an operational, though possibly narrow, definition of
equilibrium, which we will, nevertheless, use within this paper.
As explained below, its numerical implementation will provide us with a
rigorous analysis tool.

\section{\label{sec:three}Testing historical data}

\begin{figure}
\center\includegraphics[angle=0,width=100mm]{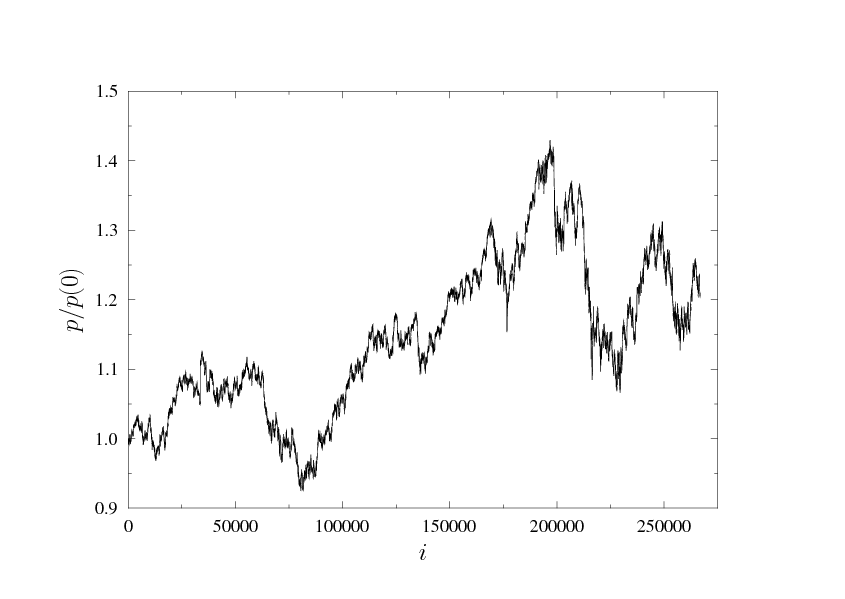}
\caption{\label{fig:price}Relative price $p/p(0)$ of the NASDAQ~100 index versus the quote
time counter $i$ for the time interval from 2005-Aug-26 to 2008-Aug-25.}
\end{figure}

The data set for our analysis is the price time series of the NASDAQ~100 stock index.
We use data from 2005-Aug-26 to 2008-Aug-25.
The size of the set is $N=266906$ \cite{Finam:2008}. 
Within normal trading hours (Monday through Friday, 09:30h to 16:00h) this translates to
an average of about $1.14$ minutes between trades.
The price time series is shown in Fig.~\ref{fig:price}.
\begin{figure}
\center\includegraphics[angle=0,width=100mm]{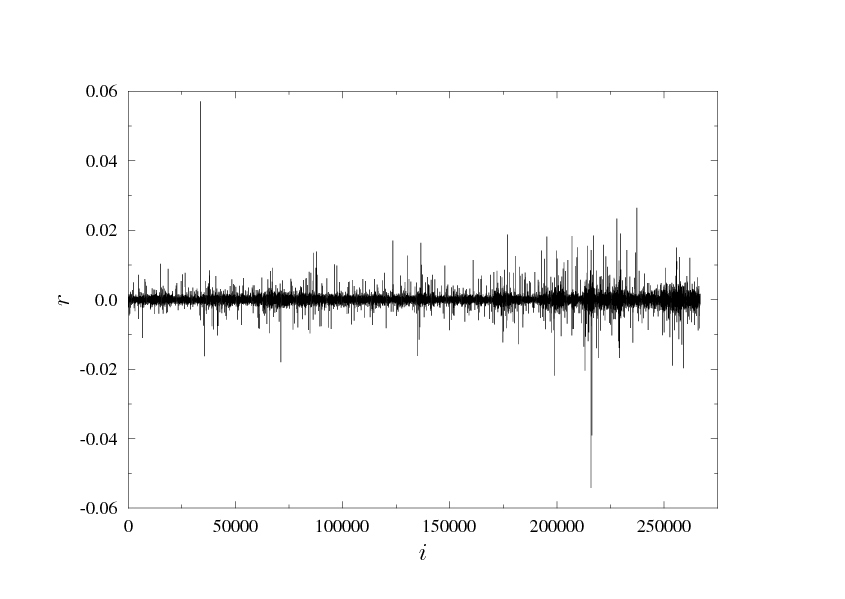}
\caption{\label{fig:return}Returns (\protect\ref{return}) corresponding to the price time
series of Fig.~\protect\ref{fig:price}.}
\end{figure}
The counter $i$ starts from $i=0$ and is incremented by one every time a new quote emerges,
$i=0,1\ldots m=N-1=266905$.
A commonly used derived measure are the returns
\begin{equation}
r(i) = \log(p(i)/p(i-1)) \quad {\rm for} \quad i=1\dots m\,,
\label{return}\end{equation}
where $\log$ here means the natural logarithm.
The returns corresponding to Fig.~\ref{fig:price}, are displayed in Fig.~\ref{fig:return}.
The the original price time series $p(i)$ can be easily
reconstructed from $r(i)$ by way of recursion, $p(i)=p(i-1)\exp{r(i)}$,
given the initial condition, which is $p(0)=1565.87${\sc USD}.

Obviously, as the price is strongly changing with time, see Fig.~\ref{fig:price},
the notion of equilibrium cannot apply directly to the price,
at least not on the time scale considered.
Thus, in contrast to mainstream economic practice, we will rather test the returns
time series for equilibrium (viz. detailed balance).
A glance at Fig.~\ref{fig:return} shows that this is
a much more reasonable starting point.  
In this sense, we deviate from the colloquial use of the term equilibrium in an
economic context.

\begin{figure}
\center\includegraphics[width=100mm]{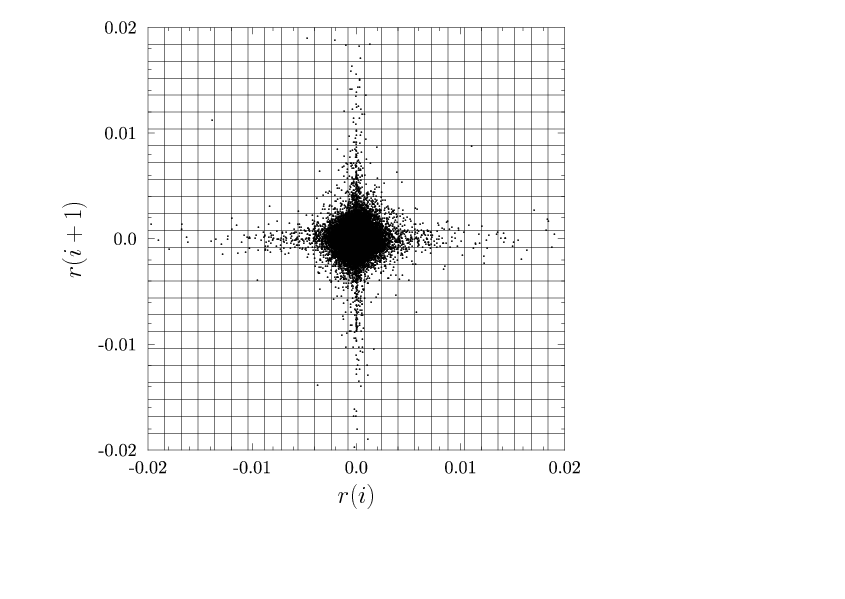}
\caption{\label{fig:P2density-1}Plot of transition events $r(i+1)\leftarrow r(i)$
extracted from the returns time series of Fig.~\protect\ref{fig:return}. Each event is shown as
a dot. The overlay shows the square bins used in the discretization of the data.}
\end{figure}

\begin{figure}
\center\includegraphics[angle=0,width=100mm]{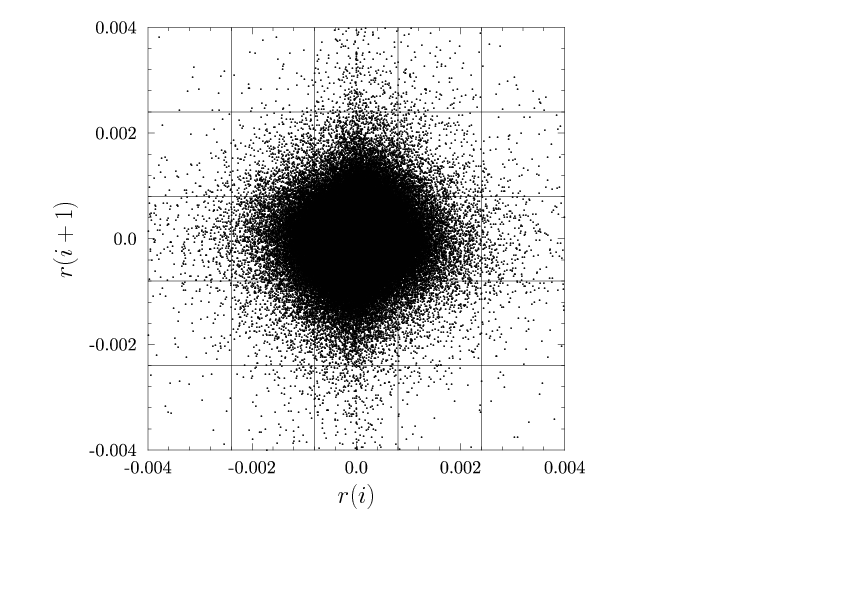}
\caption{\label{fig:P2density-2}A zoom-in towards the center of Fig.~\protect\ref{fig:P2density-1}.}
\end{figure}

The extraction of the transition probability density
$W(r^\prime\leftarrow r)$ in (\ref{Dbalance}) from the historical data set Fig.~\ref{fig:return}
is illustrated in Fig.~\ref{fig:P2density-1}.
The pair of a return event $r(i)$ and its immediate successor
$r(i+1)$ gives rise to a dot in one of the square bins of Fig.~\ref{fig:P2density-1}.
The number of dot counts, in a particular square bin, then is an estimator for
the transition probability density $W(r^\prime\leftarrow r)$ in (\ref{Dbalance}), up to
a normalization factor.
Within intervals $r\in[-0.02,+0.02]$, for each axis, we chose a discretization of $\Delta r=0.0016$,
which translates into $N=25$ bins in each direction, or a total of $625$ square bins.
The data are very heavily peaked at the center, leading to saturation near the origin.
The counts displayed in Fig.~\ref{fig:P2density-1} accommodate all but 7 of the returns events
of the data set.
Fig.~\ref{fig:P2density-2} is just a detail of Fig.~\ref{fig:P2density-1} at a smaller scale.

It is convenient to adapt our notation to the discretization. Thus let $x,y\in\mathbb{N}$
denote discrete bin counters, $x,y=1 \ldots N$, so each square bin is labeled by a pair $(x,y)$.
Thus $W(x,y)$ is the number of counts in square bin $(x,y)$. Let $w(x)$ be the discretized
version of the probability distribution function $w(r)$ as it appears in (\ref{Dbalance}).
The aim of our analysis is to find a discretized probability density $w(x)$ such that
\begin{equation}
S[w] = \frac{1}{K}\sum^\prime_{1 \le x < y \le N}
\left[\frac{W(x,y)w(y)-W(y,x)w(x)}{W(x,y)w(y)+W(y,x)w(x)}\right]^2
\label{Saction}\end{equation}
is a minimum. We refer to $S[w]$ an the action functional.
The prime ${}^\prime$ on the sum is meant to indicate the restriction $W(x,y)w(y)+W(y,x)w(x)>0$. 
Since $W(x,y)w(y) \ge 0$ and $W(y,x)w(x) \ge 0$ both apply, the condition\\
$W(x,y)w(y)+W(y,x)w(x)=0$ would
imply that also $W(x,y)w(y)-W(y,x)w(x)=0$, thus satisfying detailed balance trivially, in a bin. 
Hence, those terms may be dropped from the sum in (\ref{Saction}) without harm to the desired
utility of $S[w]$ as an indicator for detailed balance.
The normalization constant $K$ in (\ref{Saction}) is simply the number of (non-zero) terms
contributing to the sum. It can reach the maximal value of $K=N(N-1)/2$.
Finally, for real numbers $a \ge 0, b \ge 0$ with $a+b>0$ it is easy to show that
$-1 \le (a-b)/(a+b) \le +1$. Thus we see that the action (\ref{Saction}) is in the
range $0 \le S[w] \le 1$. While $S[w]=0$ indicates exact detailed balance
\begin{equation}
W(x,y) w(y) = W(y,x) w(x) \,,
\label{Sbalance}\end{equation}
$S[w]=1$ means that (\ref{Sbalance}) is maximally violated.
Thus, in some sense, the action $S[w]$ is a an indicator for a market being in the range
of `completely efficient' ($S[w]=0$) and its opposite ($S[w]=1$) being in a state `off
equilibrium', including a (self-organized) critical, or a random, state.

The raw data for the transition density $W(x,y)$ are merely counts of historical events,
see Figs.~\ref{fig:P2density-1} and \ref{fig:P2density-2}. Those counts depend on the
size of the data set and thus require some normalization before they can be interpreted as a
probability density. Define the normalization constants
\begin{equation}
C(y)=\sum_{x}W(x,y)\,.
\label{Cnorm}\end{equation}
Our choice of binning the data ensures that $C(y)>0$ for all $y$. Thus define
\begin{eqnarray}
\hat{W}(x,y) &=& W(x,y)/C(y) \label{hatW}\\
\hat{w}(y) &=& C(y)w(y) \label{hatw}
\end{eqnarray}
Clearly, detailed balance is preserved
\begin{equation}
\hat{W}(x,y) \hat{w}(y) = \hat{W}(y,x) \hat{w}(x) \,,
\label{Wbalance}\end{equation}
and the action (\ref{Saction}) remains invariant.
However, through (\ref{hatw}), the base probability density $w$ is
sensitive to the normalization of $W$. Noting that
\begin{equation}
\sum_{x}\hat{W}(x,y) = 1\,,
\label{hatW1}\end{equation}
we see from (\ref{Wbalance}) that
\begin{equation}
\hat{w}(y) = \sum_{x} \hat{W}(y,x) \hat{w}(x) \,.
\label{Wfix}\end{equation}
Hence the distribution $\hat{w}$ is a fixed point of $\hat{W}$,
the very meaning of equilibrium.
The distribution $\hat{w}$ thus is the goal of our analysis.

\section{\label{sec:four}Numerical implementation and results}

We compute an optimal solution $w(x)$ to $S[w]=\min$ by way of simulated
annealing \cite{Kirkpatrick13051983}. In a nutshell this, standard, technique \cite{Press:2007}
revolves around the partition function
\begin{equation}
Z = \sum_{[w]}\,e^{-\beta S[w]}\,,
\label{Zpart}\end{equation}
where the sum is over all possible probability distribution functions $w(x), x=1\ldots N$,
called configurations $[w]$ in this context. There are two main ingredients to this strategy:
First, at any given `temperature' $T=\beta^{-1}$ one employs a Metropolis
algorithm \cite{metropolis1953} to achieve equilibrium\footnote{In this context it is
just a technical tool used for simulated annealing. It has nothing to do with our, motivational,
mention of the Metropolis algorithm in Sect.~\protect\ref{sec:two}.},
resulting in configurations
$[w]$ drawn from the probability distribution function $e^{-\beta S[w]}$.
Second, the temperature is gradually decreased according to a chosen annealing
schedule in the range $\beta_1 < \beta_2$, very slowly cooling down the system.
If carefully conducted, the system will settle into a state of a global
minimum $S[w]=\min$.
\begin{figure}
\center\includegraphics[angle=0,width=100mm]{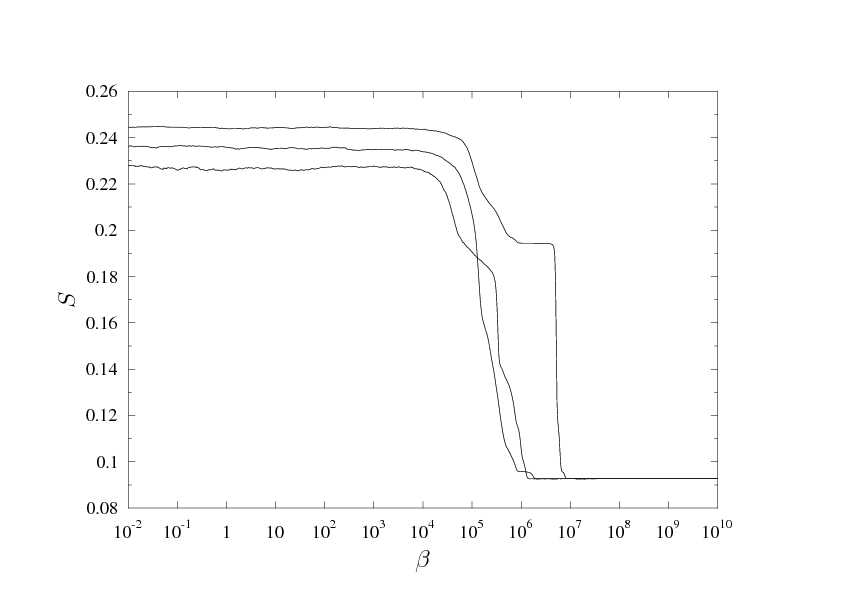}
\caption{\label{fig:annealA}Annealing histories of the action $S[w]$ defined
in (\protect\ref{Saction}) versus $\beta$, the inverse `temperature'.
The three graphs correspond to different random starts.}
\end{figure}
\begin{figure}
\center\includegraphics[angle=0,width=100mm]{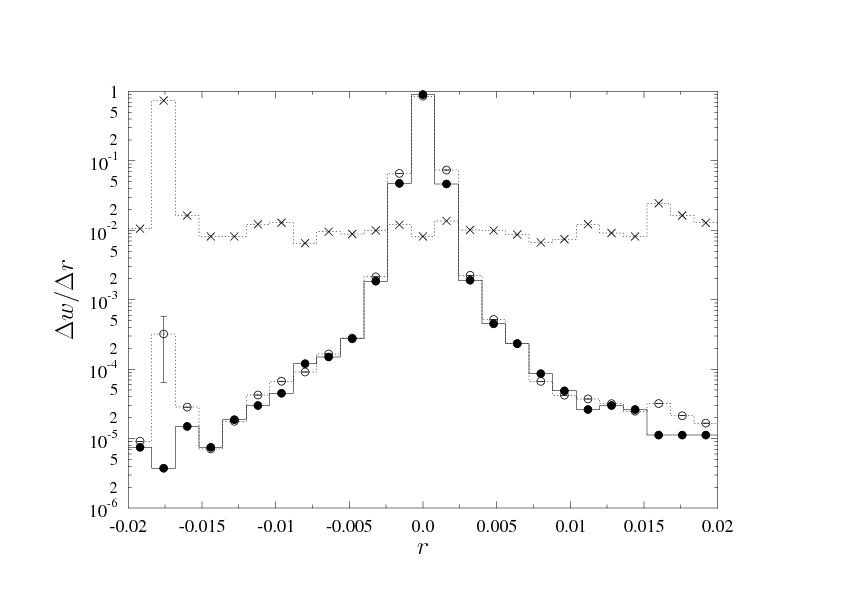}
\caption{\label{fig:returndA}The returns distribution extracted from the historical data set,
see Fig.~\protect\ref{fig:return}, is shown as a solid line histogram with filled circles
$\bullet$ as plot symbols. The distribution obtained from
minimizing the detailed balance action (\protect\ref{Saction}), using the normalized
transition density $\hat{W}$, see (\ref{hatW1}), is displayed as a
dotted line histogram with open circles $\circ$ as plot symbols. The remaining histogram,
marked with $\times$ plot symbols, corresponds to runs with the unnormalized
transition density $W$.}
\end{figure}

For the Metropolis steps, each at $\beta={\rm const}$, we employ $1600$ sweeps on a
current configuration $[w]$. A sweep consists of successively updating all $N$ components,
$w(x)$, one at a time, via
\begin{equation}
w^\prime(x) = w(x)(1+t\epsilon)
\label{sweep}\end{equation}
where $t\in[-1,+1]$ is a uniform random number, and our choice is $\epsilon=0.001$.
The configuration is then normalized, $\sum_{x}w^\prime(x)=1$. Next, the trial $w^\prime(x)$
is accepted or rejected according to the Metropolis prescription.
The chosen annealing schedule is given by 
\begin{equation}
\beta=\beta_1 e^{bj} \quad j=0\ldots n\,,
\label{anneal}\end{equation}
with $\beta_1=10^{-2}$ and $n=800$. Setting $b=0.0345388$ then gives a terminal
$\beta_2=10^{10}$, at $j=n$.  

On a cluster with 48 processors in parallel numerous runs gave very consistent results.
For each run we have chosen 48 random start configurations $[w]$ subject to\\
\mbox{$w(x)>0$} and $\sum_{x=1}^{N}w(x)=1$, where the $w(x)$ were random numbers drawn from
a uniform distribution.
Samples of the annealing history of such a run are displayed in Fig.~\ref{fig:annealA}.
The three histories correspond to the
largest, smallest, and median initial actions $S[w]$, at $\beta_1=10^{-2}$.
Invariably, all runs settle in at $S=0.0926398$ indicating a distinct global minimum of 
the action (\ref{Saction}). 
The corresponding, optimal, returns distribution using the normalized transition
density (\ref{hatW1}) is displayed in Fig.~\ref{fig:returndA}.
The, statistical, errors on the latter stem from $48$ simulated annealing starts.
With the exception of one data point the errors are invisible because of their smallness.
Given the action's theoretical range $0 \le S \le 1$, a value of $S\approx 0.1$
is close to the bottom of the scale. The visual impression given by the computed returns
in Fig.~\ref{fig:returndA} is consistent with this result.
However, the $\log$ scale of Fig.~\ref{fig:returndA} obscures the fact that the computed
returns are off by a factor of $\approx 2$ compared to the historical data set
($2.2$ at $r=-0.012$ and $1.6$ at $r=+0.016$).
 
\begin{figure}
\center\includegraphics[angle=0,width=100mm]{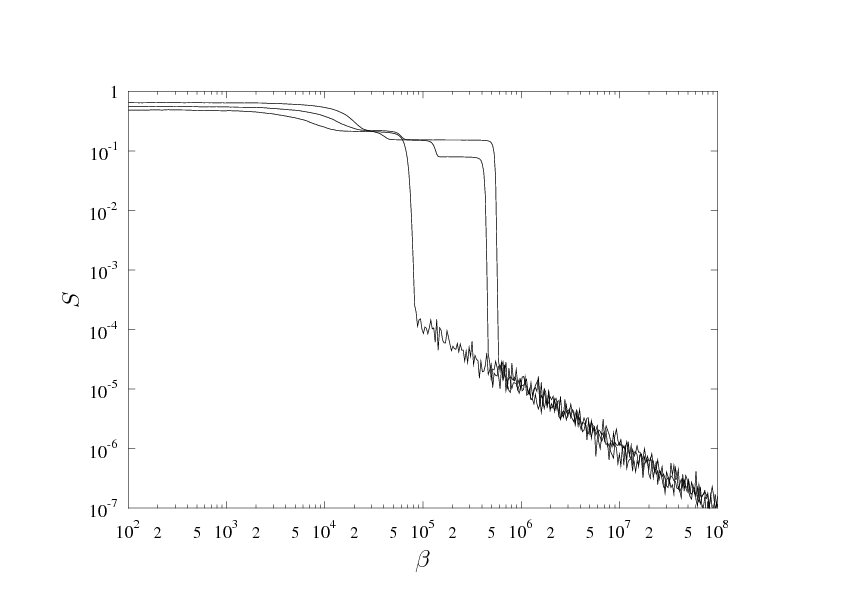}
\caption{\label{fig:annealP}Typical annealing histories, like in Fig.~\protect\ref{fig:annealA},
but using the Metropolis transition probability (\protect\ref{Dmetro}).}
\end{figure}

\begin{figure}
\center\includegraphics[angle=0,width=100mm]{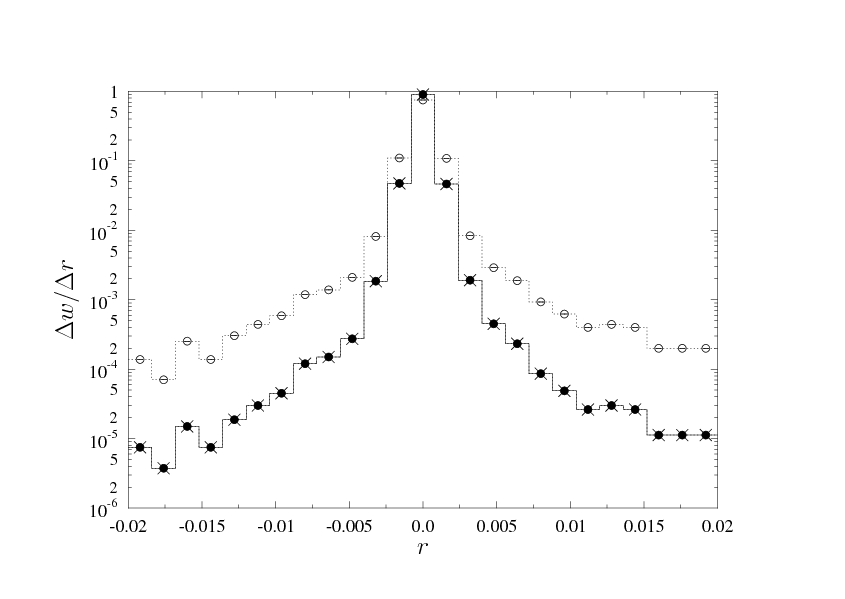}
\caption{\label{fig:returndP}Comparison of returns distributions, as described
in Fig.~\protect\ref{fig:returndA}, but using the Metropolis
transition probability (\protect\ref{Dmetro}).}
\end{figure}

In order to evaluate this situation (and also to gain confidence in the code) we have
replaced the empirical transition probabilities
$W(x,y)$ with those used in the Metropolis algorithm. They are constructed from the base
distribution $w$ as
\begin{equation}
W(x,y) = \min\left(\frac{w(x)}{w(y)},1\right)\,.
\label{Dmetro}\end{equation}
This choice satisfies detailed balance (\ref{Dbalance}) exactly.
We take $w(x)$ to be the empirical returns distribution, as displayed in Fig.~\ref{fig:returndA}
as filled circles plot symbols.
The numerical framework (binning, annealing schedule, random starts, etc)
was kept exactly as described before.
Typical annealing histories, again for three random starts, are shown in Fig.~\ref{fig:annealP}.
For all runs the algorithm finds the minimum of the action with convincing ease. We also see that
its numerical value, $S\approx 10^{-7}$ (eventually reaching machine precision) is substantially
less than $S\approx 0.1$ found with
the empirical transition probability distribution, see Fig.~\ref{fig:annealA}.
The optimal returns distributions are shown in Fig.~\ref{fig:returndP}.
The distribution $w$ computed with the unnormalized transition density (\ref{Dmetro}),
marked with $\times$ plot symbols, matches the input returns distribution exactly, which
it should, thus validating the algorithm.
The distribution $\hat{w}$ computed with the normalized transition density $\hat{W}$,
marked with $\circ$ plot symbols in Fig.~\ref{fig:returndP}, has the expected features 
(strong peak, fat tails) and is in this respect similar to the corresponding distribution
based on the empirical transition density, see Fig.~\ref{fig:returndA}.
Despite their likeness, the indicators for detailed balance are significantly different,
$\log S \approx -2.4$ and  $\log S \approx -7.0$ respectively, thus suggesting that
that the underlying stochastic dynamics may not have much in common.

\begin{figure}
\center\includegraphics[angle=0,width=100mm]{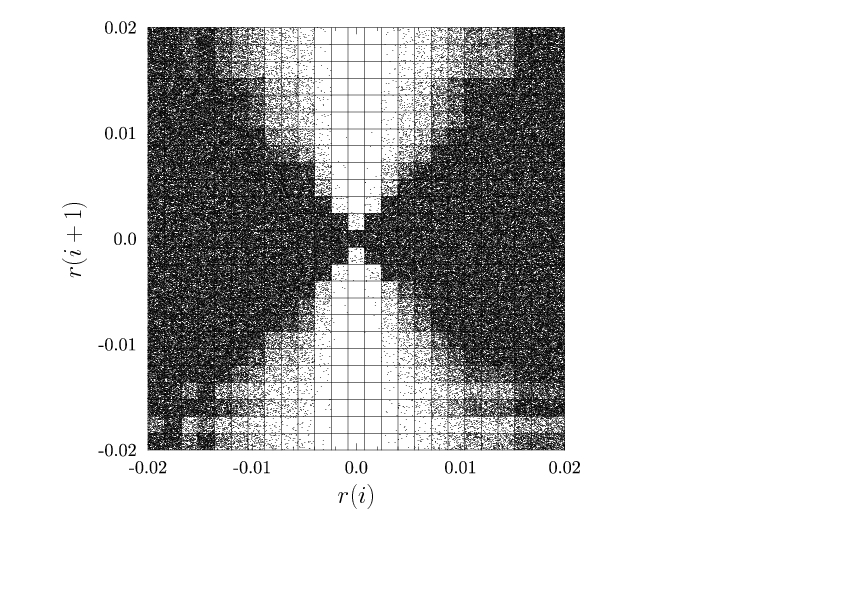}
\caption{\label{fig:P2density}Transition probability density distribution $W(x,y)$
for the Metropolis choice (\protect\ref{Dmetro}). Data binning is done like
in Fig.~\protect\ref{fig:P2density-1}.}
\end{figure}

In order to corroborate this statement, we present in Fig.~\ref{fig:P2density} the transition
density distribution $W(x,y)$ of the Metropolis distribution (\ref{Dmetro}).
This distribution has to be juxtaposed with the historical one of Fig.~\ref{fig:P2density-1}.
Clearly, their patterns could not be more different. We take this as another indication that the
historical returns (let alone the price) time series is not following statistics consistent
with an equilibrium described by detailed balance.

\begin{figure}
\center\includegraphics[angle=0,width=100mm]{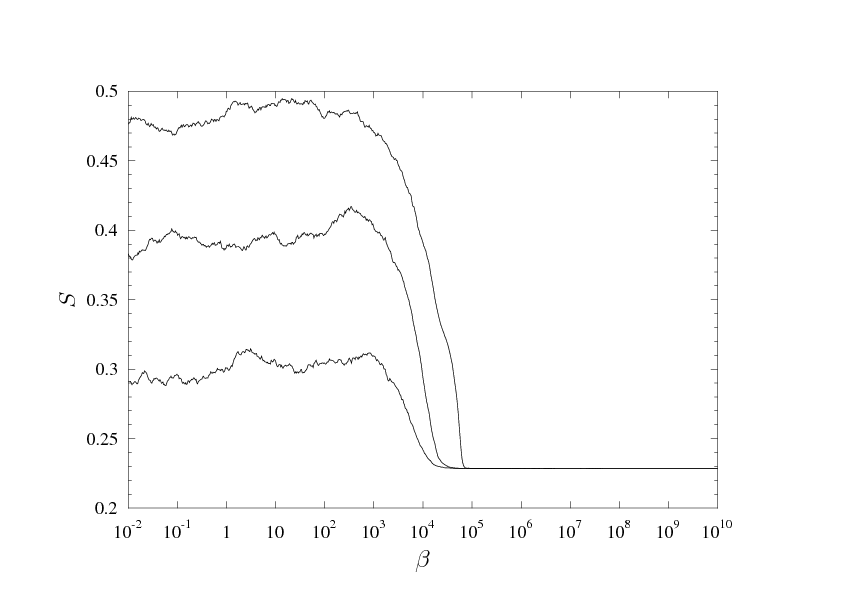}
\caption{\label{fig:annealR}Typical annealing histories, like in Fig.~\protect\ref{fig:annealA},
but using a uniform random distribution for the transition probability $W(x,y)$}.
\end{figure}

As an additional control feature, we have also computed a returns distributions obtained
from replacing the transition probability $W(x,y)$ in (\ref{Saction}) with
uniform random numbers.
The corresponding annealing history is displayed in Fig.~\ref{fig:annealR},
\begin{figure}
\center\includegraphics[angle=0,width=100mm]{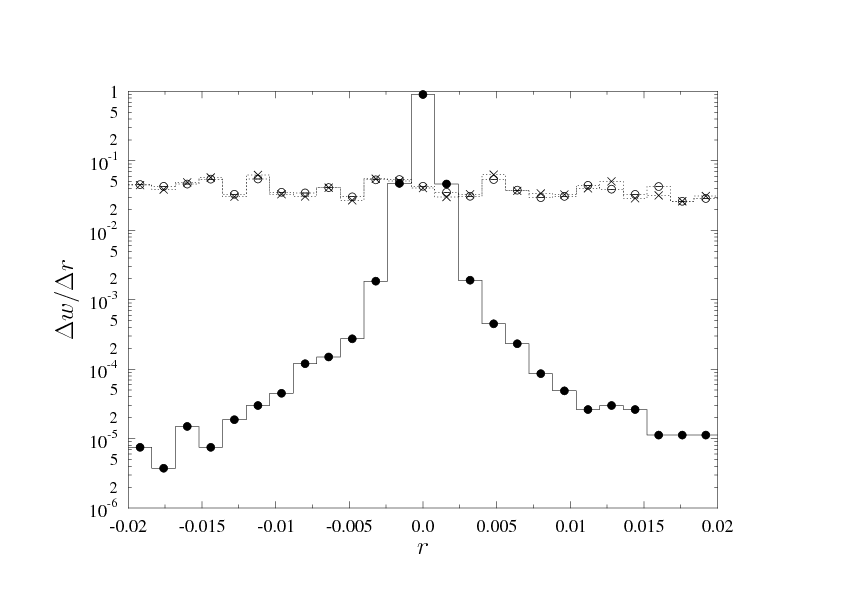}
\caption{\label{fig:returndR}Comparison of returns distributions, as described
in Fig.~\protect\ref{fig:returndA}, but using a random uniform
transition probability (\protect\ref{Dmetro}).}
\end{figure}
Again, we find a unique minimum of the action, here at $S=0.228325$, or $\log S \approx -1.5$.
The resulting returns densities, with and without normalization, are shown
in Fig.~\ref{fig:returndR}. They are essentially flat, nowhere resembling empirical features.
Nevertheless, the detailed balance action signal $\log S \approx -2.4$ is not far from the
signal of the random case $-1.5$, certainly a big distance from $-16.0$, the case of
exact (machine precision) detailed balance.

The picture emerging from our analysis thus is that the visual impressions of returns
distributions, with all its stylized features, is not a reliable indicator of a market
in equilibrium (as defined within the confines of this paper).
However, the proposed measure $S$ defined in (\ref{Saction}),
more appropriately $\log S$, appears to be very sensitive
to this feature.

\section{\label{sec:conclusion}Summary and conclusion}

We have tested a historical financial price time series for detailed balance, which is
a statistical property that, if present, would indicate a market condition known as
equilibrium in neo-classical economic theory.
Rather than using the prices directly, the test is devised around the time series of
returns. Based on the empirical transition probability of returns, between subsequent
trade signals, we define a functional $S[w]$ of the returns probability distribution $w$ 
and find its minimum $S$ by way of simulated annealing.
Within its range $0 \le S \le 1$, where $S=0$ means exact detailed balance, we find
$\log S = -2.4$ for the historical time series, whereas a control run with a randomly
generated transition probability of returns yield $\log S = -1.5$. 
The visual impression of the computed returns distribution (at $-2.4$) has features
of a realistic one, whereas the control distribution (at $-1.5$) has not. 

Our conclusion thus is that the historical returns distribution does have elements
of equilibrium, as measured by $\log S$, but is certainly far from being numerically
exact, which would be indicated by $\log S \approx -16.0$, a typical machine precision.
Although we think that the proposed test may be a useful analysis tool testing
for `equilibrium' in time series, it is also clear that it must be applied to a
larger variety historical data sets to validate its utility.
In particular, it would be interesting to see it applied to high-frequency financial
data.

%\bibliography{./qfcite}

\end{document}